\documentclass[prl,twocolumn,superscriptaddress,nofootinbib]{revtex4}

\usepackage{color}
\usepackage{graphicx,bm,times}
\usepackage{lipsum}
\usepackage{amsmath}
\usepackage{gensymb}
\usepackage{color,soul, mathrsfs}
\usepackage[]{natbib}
\usepackage{mathtools}
\usepackage[pdftex,colorlinks=true,linkcolor=blue,citecolor=blue,urlcolor=blue]{hyperref}
\usepackage{amsmath, amsthm, amssymb}
\usepackage[caption=false]{subfig}
\usepackage{comment}
\usepackage{tikz}
\usepackage{bbold}
\usepackage{balance}
\usepackage{siunitx}
\usepackage{braket}
\usepackage{filecontents}

\begin{document}

\title{Shot-Noise Limited Homodyne Detection for \si{\mega\hertz} Quantum Light Characterisation in the \SI{2}{\micro\meter} Band}

\author{Jake Biele}
\email{jb12101@bristol.ac.uk}
\affiliation{Quantum Engineering Technology Labs, H. H. Wills Physics Laboratory and Department of Electrical \& Electronic Engineering, University of Bristol, BS8 1FD, United Kingdom.}
\affiliation{Quantum Engineering Centre for Doctoral Training, H. H. Wills Physics Laboratory and Department of Electrical \& Electronic Engineering, University of Bristol, Tyndall Avenue, BS8 1FD, United Kingdom.}
\author{Joel F. Tasker}
\affiliation{Quantum Engineering Technology Labs, H. H. Wills Physics Laboratory and Department of Electrical \& Electronic Engineering, University of Bristol, BS8 1FD, United Kingdom.}
\author{Joshua W. Silverstone}
\affiliation{Quantum Engineering Technology Labs, H. H. Wills Physics Laboratory and Department of Electrical \& Electronic Engineering, University of Bristol, BS8 1FD, United Kingdom.}
\author{Jonathan C. F. Matthews}
\email{jonathan.matthews@bristol.ac.uk}
\affiliation{Quantum Engineering Technology Labs, H. H. Wills Physics Laboratory and Department of Electrical \& Electronic Engineering, University of Bristol, BS8 1FD, United Kingdom.}

\begin{abstract}
Characterising quantum states of light in the \SI{2}{\micro\meter} band requires high-performance shot-noise limited detectors. Here, we present the characterisation of a homodyne detector that we use to observe vacuum shot-noise via homodyne measurement with a \SI{2.07}{\micro\meter} pulsed mode-locked laser. The device is designed primarily for pulsed illumination. It has a \SI{3}{\decibel} bandwidth of \SI{13.2}{\mega\hertz}, total conversion efficiency of 58\% at \SI{2.07}{\micro\meter} and a common-mode rejection ratio of \SI{48}{\decibel} at \SI{39.5}{\mega\hertz}. The detector begins to saturate at \SI{1.8}{\milli\watt} with \SI{9}{\decibel} of shot-noise clearance at \SI{5}{\mega\hertz}. This demonstration enables the characterisation of megahertz-quantum optical behaviour in the \SI{2}{\micro\meter} band.
\end{abstract}

\date{\today}

\maketitle

\section{Introduction}

In recent years a need has emerged for low noise detection of weak optical fields for the characterisation of quantum states of light in the short and mid-infrared spectral region ($\geq$ \SI{2}{\micro\meter}). This has largely been driven by short-wave infrared (SWIR) and mid-infrared (MIR) quantum state preparation under development for photonic quantum technologies. Key examples include quantum-enhanced sensing applications in the fingerprint regime~\cite{Barsotti2018,Yap2019Squeezedm}, free-space communications and satellite-based quantum key distribution~\cite{Temporao2008FeasibilityMID-INFRARED,Dequal2021FeasibilityDistribution}, which better exploit the atmospheric transparency window, and as a route to avoid nonlinear loss at the C-band in silicon quantum photonics for photonic quantum computing~\cite{Rosenfeld2020Mid-infraredSilicon}.

The power of shot-noise limited homodyne detection in measuring and characterising quantum light fields~\cite{Lvovsky2009Continuous-variableTomography} has meant that it has become ubiquitous in quantum technology demonstrations based on visible and near-IR wavelengths for a range of different applications~\cite{Fuwa2015ExperimentalMeasurements,Gross2011AtomicStates,Jouguet2013ExperimentalDistribution,Breitenbach1997, Ourjoumtsev2007GenerationStates,Gabriel2010AStates,Millot2016Frequency-agileSpectroscopy,Yang2021AChip,Furusawa1998UnconditionalTeleportation} and as a result, is now being integrated to enable miniaturisation and technology scale-up~\cite{Tasker2021SiliconLight,Bruynsteen2021IntegratedMeasurements}. The majority of shot-noise limited homodyne detection has so far been limited to the visible and telecommunications band for two main reasons: (1) high-efficiency photodiodes have historically only been available in these bands; and (2) key optical infrastructure such as low noise lasers and passive components has been lacking outside these regions, limiting the range of possible experiments.

Balanced detection, a classical technique similar to homodyne detection, has already been investigated in the MIR for classical applications such as frequency-modulation spectroscopy~\cite{Carlisle1989Tunable-diode-laserDetection}, difference-frequency laser spectroscopy~\cite{Chen1998Difference-frequencyConstituent}, balanced radiometric detection~\cite{Sonnenfroh2001ApplicationLasers}, and Doppler-free spectroscopy~\cite{Bartalini2009Doppler-freem}. Shot-noise limited homodyne detection for quantum state measurement requires more stringent electrical noise suppression than balanced detection and is, therefore, harder to engineer. Using a pre-amplification scheme for each photodetector and a post-processed photocurrent subtraction, Mansell \textit{et al.} and Yap \textit{et al.} both employ extended InGaAs photodiodes in a homodyne configuration to measure squeezing at \SI{1984}{\nano\meter} in the audio band~\cite{Mansell2018ObservationRegion,Yap2019Squeezedmb}. In both experiments, the measurement bandwidths are restricted to $\leq$~\SI{500}{\kilo\hertz} by the need for high gain amplification required to provide shot-noise clearance above excessive electrical noise. Using the same post-processing configuration, Gabbrielli \textit{et al.}~\cite{Gabbrielli2021Mid-infraredCharacterization} recently demonstrated shot-noise limited homodyne detection in the MIR (around \SI{4.5}{\micro\meter}) enabling the first characterisation of quantum noise with a quantum cascade laser (QCL) source. Gabbrielli \textit{et al.} were able to overcome the low conversion efficiency of state-of-the-art \SI{4.5}{\micro\meter} HgCdTe detectors to measure shot-noise with a total efficiency of 38\% and maximum clearance of \SI{7}{\decibel}.

Here, we move beyond the audio band and employ extended InGaAs PIN photodiodes (Laser Components IG26X250S4i) integrated into a single device capable of megahertz-speed shot-noise limited homodyne detection in the \SI{2}{\micro\meter} band with 58\% efficiency. This is already sufficient to use with squeezed light generation to construct optical sensors with sensitivity~\cite{Grangier1987Squeezed-lightInterferometer} and precision~\cite{Atkinson2021QuantumLight} beyond the classical limit. This efficiency is also greater than the 50\% threshold required to measure negative-valued Wigner functions, which is a clear signature of non-classical behaviour for non-Gaussian quantum states~\cite{Biagi2021ExperimentalInequalities}. We test this detector by using it to perform a homodyne measurement of vacuum shot-noise via a \SI{2.07}{\micro\meter} pulsed mode-locked fibre laser, proving the device capable of quantum noise measurement. We characterise the efficiency, bandwidth, shot-noise clearance (SNC), and common-mode rejection ratio (CMRR) of the device; these are each important considerations for using the detector to perform future squeezed light detection at \SI{2.07}{\micro\meter}.

\section{Homodyne Theory \& Design Considerations}

Homodyne detection measures field quadratures by mixing the target field at a 50:50 beamsplitter with a bright optical field  referred to as the local oscillator (LO) approximated by $\alpha(t)=|\alpha(t)|e^{i\phi}$. The two resulting beams are converted to photocurrents by high-efficiency photodiodes, and the difference between these photocurrents is amplified to give the homodyne signal. This RF signal is proportional to the target field's quadrature, with the quadrature-phase determined directly by the phase of the LO. Low noise amplification electronics are crucial for the accurate and efficient assessment of quantum noise in the target light field. Furthermore, the LO and the target field must be mutually phase stable and have a similar optical frequency. The homodyne detector reported here uses a transimpedance amplifier (TIA) to amplify the subtracted current into an output voltage~\cite{Graeme1996PhotodiodeSolutions}. Employing TIA, over passive conversion, helps enable a greater measurement bandwidth with favourable noise statistics~\cite{Sackinger2017AnalysisReceivers}. 

To test a homodyne detector's efficacy of quantum noise measurement, we perform homodyne detection on the vacuum state. The target quadrature is ${\hat{q}_{\mathrm{vac}}(t)=\hat{a}_{\mathrm{vac}}(t)e^{i\phi}+\hat{a}_{\mathrm{vac}}^\dagger(t)e^{-i\phi}}$. Since measurements of vacuum quadrature are inherently phase invariant, we omit phase dependence from the following standard analysis. Under the approximation that the pulse width is much shorter than the time resolution of the electronics, the output subtraction current is amplified into a voltage given by~\cite{Kumar2012VersatileTomography}:
\begin{equation}\label{equ:1}
\begin{split}
    \hat{V}(t) &=R_f\hat{i}(t)\\
     &=R_f\Big[\eta_\mathrm{coup}\mathcal{R}\alpha_{\mathrm{LO}}r(t)\hat{Q} + \hat{i}_e(t)\Big] 
\end{split}
\end{equation}
Here, $\hat{i}_e(t)$ is the detector dark current, $r(t)$ is the unit impulse response function of the detector, $R_f$ is the feedback gain resistor of the TIA circuit, $\eta_\mathrm{coup}$ is the coupling efficiency and $\mathcal{R}$ is the photodiode responsivity. $\hat{Q}$ is the normalised quadrature operator corresponding to the signal mode defined by the shape of the LO pulse and is given by ${\hat{Q}=\alpha_{\mathrm{LO}}^{-1}\int_{-\infty}^{\infty}\alpha(t)\hat{q}_{\mathrm{vac}}(t)dt}$, where ${\alpha_{\mathrm{LO}}=\sqrt{\int_{-\infty}^{\infty}|\alpha(t)|^2dt}}$ is a normalisation coefficient~\cite{Kumar2012VersatileTomography}. Equation~\ref{equ:1} holds under the assumptions that the total efficiency is equal for both photodiodes, the 50:50 homodyne beamsplitter is ideal and the LO is bright in comparison to the vacuum fluctuations. In reality, each photodiode will have a slightly different responsivity and coupling coefficient, and the beamsplitter will not be perfectly 50:50. These imperfections can be minimised by fine-tuning the coupling of each photodiode to ensure the detector remains balanced. A full analysis of the imperfect case can be found in references~\cite{Kumar2012VersatileTomography,Almeida2020ImpactNoise}.

We use an electronic spectrum analyser (SA) to measure the power spectral density (PSD) of the voltage signal about frequency $\Omega$ given by \cite{Atkinson2021QuantumLight}:
\begin{equation}
    \langle \hat{p}_\Omega \rangle = \frac{2}{R_\mathrm{imp}}\Bigg\langle\Bigg|\int_{\Omega-B/2}^{\Omega-B/2} \tilde{\hat{V}}(\omega) d\omega\Bigg|^2\Bigg\rangle
\end{equation}
where $\omega$ is the radio frequency, $R_\mathrm{imp}$ is the analysers input impedance and $B$ is the measurement's resolution bandwidth. Here, $\tilde{\bullet}$ denotes the Fourier transform. Using the Fourier transform of Eq.~\ref{equ:1}, we take the expectation with respect to the quantum ensemble of the signal vacuum state to give:
\begin{equation}\label{equ:psd}
    \langle \hat{p}_\Omega \rangle = \frac{2R_f^2}{R_\mathrm{imp}}\Bigg[\frac{hc}{\lambda}\eta_\mathrm{coup}^2\mathcal{R}^2\alpha_{\mathrm{LO}}^2\int_{\Omega-B/2}^{\Omega-B/2}|\tilde{r}(\omega)|^2d\omega+ |\tilde{i}_\Omega|\Bigg]
\end{equation}
A factor of photon energy, $hc/\lambda$, results from the conversion of natural to standard units. Here, $|\tilde{i}_\Omega|$ is the electronic-noise contribution integrated across the resolution bandwidth. Within the bandwidth of the detector, we assume the Fourier transform of the unit impulse response function flat such that $\int_{\Omega-B/2}^{\Omega-B/2} |\tilde{\hat{r}}(\omega)| d\omega\approx B$. By analysing how the output voltage scales with LO power, $\alpha_{LO}^2$, we asses the detectors ability to measure quadrature noise. If the detector fully rejects classical noise, the PSD will scale linearly with LO power as per Eq.~\ref{equ:psd}, if not then we expect the PSD to scale with the square of the power. We use this linear dependence to infer whether the detectors are shot-noise limited in their measurement of the vacuum and are thus suitable for sub-shot-noise quantum state measurement.

Careful TIA design is required for a flat gain spectrum, crucial for distortion-free measurements. In comparison to \SI{1550}{\nano\meter} homodyne detectors, which typically utilise InGaAs photodiodes, detection in the \SI{2}{\micro\meter} band requires extended InGaAs photodiodes that have a different material composition to increase responsivity at longer wavelengths. Due to additional strain applied in fabrication, extended photodiodes typically display a much smaller shunt resistance ($\sim\mathrm{x}10^{-3}$) allowing additional voltage supply noise to couple into the signal. Without ultra-low noise voltage supplies and additional noise decoupling considerations, it would not be possible to detect optical shot-noise at megahertz-speeds. Furthermore, extended photodiodes have a larger junction capacitance requiring careful consideration when designing for a specific detector bandwidth. Table~\ref{tab:1.1}, below, summarises the key components used and their relevant performance specifications.
\begin{table}[!ht]
\centering
\begin{tabular}{l|lll}
\multicolumn{1}{c|}{{\textbf{Component}}} & \multicolumn{1}{c}{{\textbf{Model}}} &  &  \\ \cline{1-2}
Photodiodes                                   & Laser Components: IG26X250S4i            &  &  \\ \cline{1-2}
Op-amp                                        & Analog Devices: ADA4817                  &  &  \\ \cline{1-2}
Negative Linear Regulators                    & Linear Technology: LT3094                &  &  \\ \cline{1-2}
Positive Linear Regulators                    & Linear Technology: LT3045               &  &
\vspace{2em}
\end{tabular}
\centering
\begin{tabular}{lll}
\multicolumn{1}{c|}{{\textbf{Component}}}                   & \multicolumn{1}{c|}{{\textbf{Property}}}                                       &\multicolumn{1}{c}{{\textbf{Value}}}                      \\ \hline
\multicolumn{1}{c|}{\textbf{Photodiode}} & \multicolumn{1}{l|}{Responsivity  @ \SI{2.1}{\mu m}}                & \SI{1.45}{A\per\watt}                 \\ \cline{2-3} 
\multicolumn{1}{c|}{}                            & \multicolumn{1}{l|}{Junction Capacitance ($C_\mathrm{pd}$)@ \SI{1}{\volt_b}} & \SI{9}{\pico F}                      \\ \cline{2-3} 
\multicolumn{1}{c|}{}                            & \multicolumn{1}{l|}{Shunt Resistance @ $T= $\SI{20}{\degree}}       & \SI{60}{\kilo\ohm}                   \\ \cline{2-3} 
\multicolumn{1}{c|}{}                            & \multicolumn{1}{l|}{Dark Current @ \SI{1}{\volt_b}}        & \SI{20}{\micro A}                     \\ \cline{2-3} 
\multicolumn{1}{c|}{}                            & \multicolumn{1}{l|}{Active Region Diameter}        & \SI{250}{\micro\meter}      \\ \cline{2-3}
\multicolumn{1}{c|}{}                            & \multicolumn{1}{l|}{Reverse bias}        &
\SI{2.4}{\volt}     \\ \hline
\multicolumn{1}{c|}{\textbf{Op-amp}}     & \multicolumn{1}{l|}{Gain Bandwidth Product $(A_0f_0)$}                         & \SI{410}{\mega\hertz}                \\ \cline{2-3} 
\multicolumn{1}{l|}{}                            & \multicolumn{1}{l|}{Voltage Noise @ \SI{100}{\kilo\hertz}}          & \SI{4}{\nano\volt\per\sqrt{\hertz}} \\ \cline{2-3} 
\multicolumn{1}{l|}{}                            & \multicolumn{1}{l|}{Current Noise @ \SI{100}{\kilo\hertz}}          & \SI{2.5}{\femto A\per \sqrt{\hertz}} \\ \cline{2-3} 
\multicolumn{1}{l|}{}                            & \multicolumn{1}{l|}{Input Bias Offset Current}                      & \SI{1}{\pico A}    \\ \cline{2-3} 
\multicolumn{1}{l|}{}                            & \multicolumn{1}{l|}{Input Capacitance $(C_\mathrm{oa})$}                      & \SI{1.3}{\pico F} \\ \hline
\multicolumn{1}{c|}{\textbf{Circuitry}}     & \multicolumn{1}{l|}{Gain Resistor $(R_f)$}                         & \SI{3.9}{\kilo\ohm}                \\ \cline{2-3} 
\multicolumn{1}{l|}{}                            & \multicolumn{1}{l|}{Feedback Capacitor $(C_f)$}          & \SI{4.7}{\pico F} \\
\end{tabular}
\caption{A list of key components and their key properties including the photodiodes, op-amp and feedback circuitry.}
\label{tab:1.1}
\end{table}
\begin{figure*}
\centering
    \includegraphics[scale = 0.23]{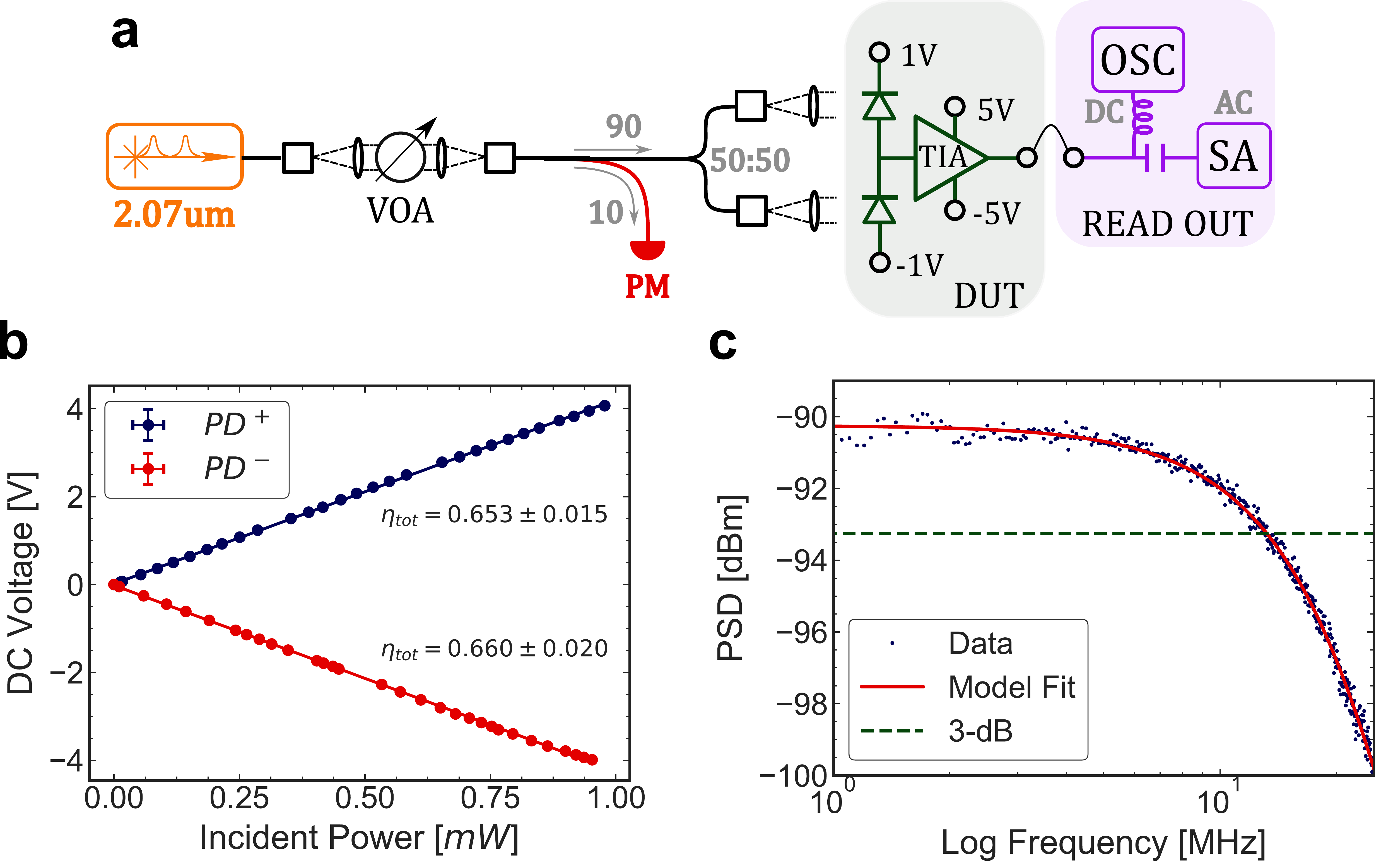}
   \caption{(a) Shows a schematic of the setup used to characterise the detectors. (b) Plots the DC response from consecutively blocking $\mathrm{PD}^+$ and $\mathrm{PD}^-$. Statistical deviation from a continuous 10-second measurement has been accounted for and total conversion efficiency has been extracted via linear regression. (c) Plots the electronic-noise corrected gain spectrum of the detector acquired with a resolution bandwidth of \SI{100}{\kilo\hertz} and video bandwidth of \SI{100}{\hertz} (blue) fit to a second-order Butterworth function (red)~Eq.~\ref{equ:butter} with $p=1.12,f^*=$~\SI{61}{\mega\hertz},~$R^2 = 0.98$.}
   \label{fig:1}
\end{figure*}

\section{Detector Characterisation}

The setup used to characterise the detector is depicted in Fig.~\ref{fig:1}(a). A pulsed \SI{2.07}{\micro\meter} fibre mode-locked laser (AdValue Photonics AP-ML-1) is used for full characterisation. The laser light passes through a free-space variable optical attenuator (VOA) and is then coupled back into SMF28 fibre. The VOA consists of a razor blade on a motorised translation stage, used to partially block the beam. The overall insertion loss of the VOA is \SI{4.1}{\decibel}. From here, the laser passes through a fibre beam splitter with an intended splitting ratio of 90:10. We characterise this splitter to have an insertion loss of \SI{0.44}{\decibel} and an actual splitting ratio of 88.7:11.3. The 11.3\% arm is detected via a power meter and is used to monitor the input power. The 88.7\% arm is split again via a 50:50 fibre splitter (Thorlabs TW2000R2F1B) with each arm coupled into the photodiodes ($\mathrm{PD}^+$ and $\mathrm{PD}^-$) of the detector. A set of lenses with anti-reflection coating (Thorlabs C037TME-D) are used to focus the beams down onto $\mathrm{PD}^+$ and $\mathrm{PD}^-$. It is important to match the path lengths to each photodiode from the fibre splitter to ensure good temporal overlap in the subtraction with imbalances in path length leading to poor common-mode rejection at higher frequencies. This is achieved by maximising the rejection of the repetition rate in the output RF signal. The RF output of the detector is split via a bias tee into its DC and AC components. The DC is coupled into an oscilloscope (Tektronix TDS2012B) and the AC component is coupled into an \SI{6}{\giga\hertz} electrical spectrum analyser (FieldFox model N9912A).

We optimise the coupling of each arm separately by temporarily blocking the other and maximising the DC signal for fixed input power. The beam must be focused onto the photodiode's active region with the optimal coverage given by a beam waist slightly less than the active area diameter. We do not focus any tighter to avoid the risk of a nonlinear response from the detector.

The total conversion efficiency, defined as the product of the coupling $\eta_{\mathrm{coup}}$ and quantum $\eta_{\mathrm{QE}}$ efficiencies, is extracted for each photodiode. This is achieved by successively blocking each photodiode and measuring the increase in DC voltage as a function of incident power. The DC voltage is given by:
\begin{equation}\label{equ:dcv}
   V_{\mathrm{DC}}=\pm\eta_\mathrm{coup}\mathcal{R}R_f\alpha_{\mathrm{
   LO}}^2 + V_{\mathrm{offset}}
\end{equation}
where $V_{\mathrm{offset}}$ is the offset voltage resulting from both the input bias offset current and output offset current of the op-amp and imbalances in the photodiode leakage current. Since the current subtraction $I_{\mathrm{PD}^+}-I_{\mathrm{PD}^-}$ is amplified, the plus-minus describes illumination of $\mathrm{PD}^+$ and $\mathrm{PD}^-$ respectively. $\mathrm{PD}^+$ and $\mathrm{PD}^-$ have total efficiencies of $65.3\%\pm1.5\%$ and $66\%\pm2\%$ respectively,~Fig.~\ref{fig:1}(b). The voltage offset is measured to be \SI{2.4}{\milli\volt} at room temperature, though this is expected to be temperature-dependent.

 \begin{figure*}
    \includegraphics[scale = 0.23]{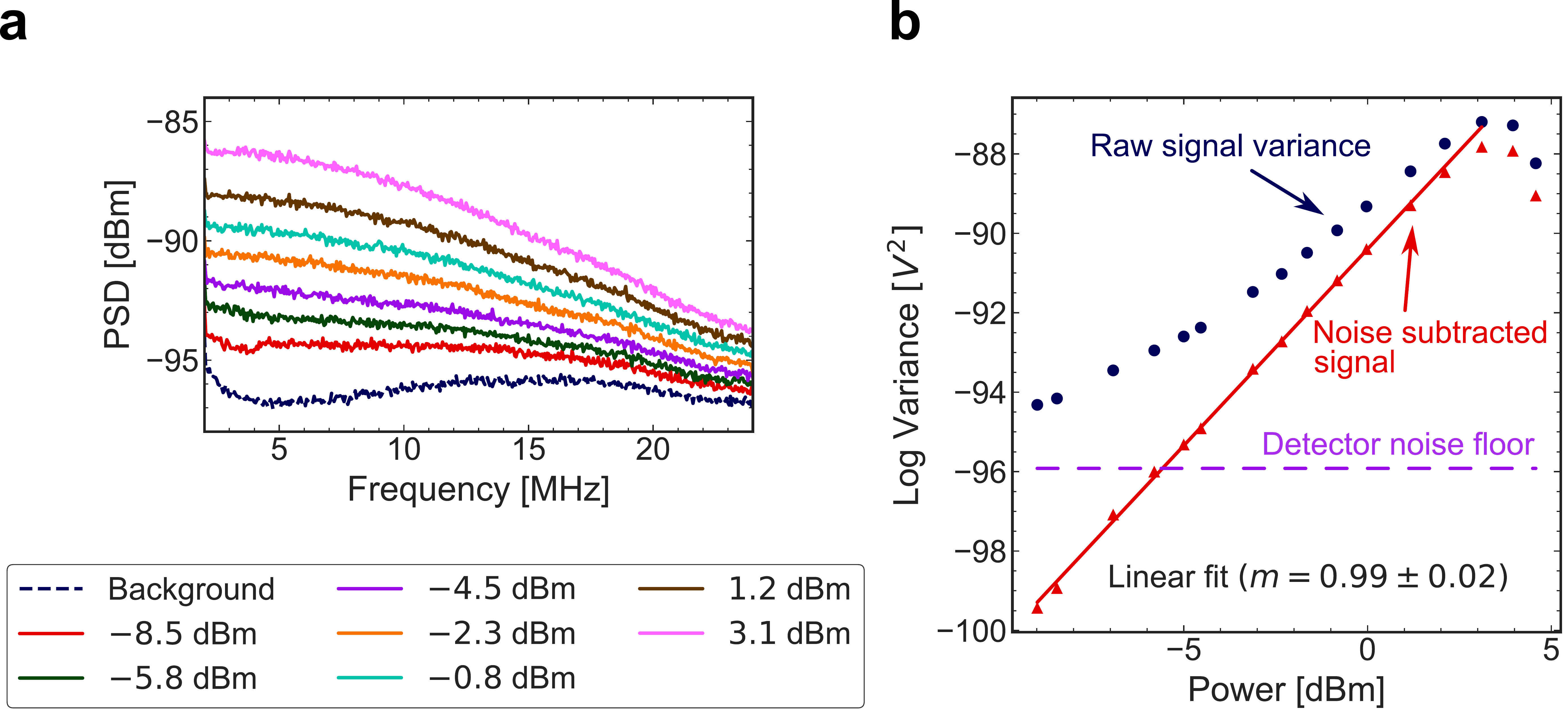}
   \caption{(a) Shows the power spectral density averaged over 30 traces for varying input power as measured in balanced configuration from pulsed illumination. The measurement resolution bandwidth is \SI{300}{\kilo\hertz}, with a video bandwidth of \SI{10}{\kilo\hertz}. A moving point average with a Gaussian kernel has been applied with a FWHM of \SI{1.5}{\mega\hertz}. (b) Plots the average noise variance over the frequency range \SI{1}{\mega\hertz}--\SI{13}{\mega\hertz} as a function of incident power (blue). Linear regression has been performed via least-squares giving a fit with $R^2=1.00$ (red).}
   \label{fig:2}
\end{figure*}
For an ideal TIA, the Fourier transform of the unit impulse response is given analytically by a second-order Butterworth function~\cite{Masalov2017}: 
\begin{equation}\label{equ:butter}
\begin{split}
    |\tilde{r}(\omega)|^2  &= \frac{1}{1+(p^2-2)\frac{\omega^2}{\omega^{*2}}+\Big(\frac{\omega^2}{\omega^{*2}}\Big)^2}\\
    p &= \Big(2\pi R_fC_f + \frac{1}{A_0f_0}\Big)\omega^*\\
    \omega^* &= \sqrt{\frac{A_0f_0}{2\pi(2C_{\mathrm{pd}}+C_f+C_{\mathrm{oa}})}}
\end{split}
\end{equation}
where $p$ and $\omega^*$ are parameterised by the components and feedback circuitry (see table~\ref{tab:1.1} for variable definitions). To produce a flat second-order Butterworth response, the feedback capacitor is chosen such that $p\approx\sqrt{2}$, at which point $\omega^*$ is then the 3-dB bandwidth of the device. Figure~\ref{fig:1}(b) plots the detector gain as a function of frequency. Fitting to the data, we obtain $p=1.12$. Fine-tuning the feedback capacitor can improve this further. Since the response is not perfectly flat, $\omega^*$ will overestimate the true bandwidth. By reading off \SI{-3}{\decibel} from the fit in Fig.~\ref{fig:1}(b) we determine  the bandwidth to be \SI{13.2}{\mega\hertz}.

For homodyne detection of the vacuum, we ensure the detector remains balanced on $V_{\mathrm{offset}}$ via the DC signal. We then measure $\langle \hat{p}_\Omega \rangle$  with the spectrum analyser for a range of input laser powers. Figure~\ref{fig:2}(a) plots the output spectrum for increasing laser powers with a resolution bandwidth of \SI{100}{\kilo\hertz}. By subtracting the electronic-noise spectra from the shot-noise response  and averaging over~\SI{1}{\mega\hertz}--\SI{13}{\mega\hertz}, we can perform a linear fit to the average voltage variance from the detector with power (Fig.~\ref{fig:2}(b)). We obtain a linear fit with an $R^2=0.99$. Note that the detector starts to display saturation above \SI{1.8}{m\watt}; these points have been omitted from the regression. The detector electronic current noise level averaged across \SI{1}{\mega\hertz}--\SI{13}{\mega\hertz} is \SI{2.79}{\pico A/\sqrt{\hertz}} with a shot-noise equivalent incident power of \SI{73}{\micro W}.

The CMRR of a homodyne detector is defined as the ratio of the signal power with one photodiode blocked compared to the balanced configuration \cite{Chi2011ADistribution}. We perform this measurement using the signal power at the repetition rate of the laser. In contrast to the definition of CMRR, we perform the blocked measurement by re-routing all optical power to one photodiode, labelled 'Addition'. When only one photodiode is used, the photocurrent is proportional to optical intensity, and thus, the PSD of the electrical signal is proportional to optical intensity squared. The addition signal from the detector contains twice the optical power (+ \SI{3}{\decibel}) compared to the subtraction and consequently the detector PSD is (+ \SI{6}{\decibel}) greater when compared to the standard definition of CMRR. We therefore correct the difference of \SI{54}{\decibel} by -\SI{6}{\decibel} to extract a true CMRR of \SI{48}{\decibel}.

\section{Discussion}

The detector quantum efficiency $\eta_{\mathrm{QE}}$ at \SI{2.1}{\micro\meter} is quoted by the manufacturer to be 86.5\%. In practice, it is difficult to decouple the true responsivity of each photodiode from the coupling efficiency $\eta_{\mathrm{coup}}$ obtained in our setup. Assuming the quoted $\eta_{\mathrm{QE}}$ is nominal, we then calculate $\eta_{\mathrm{coup}}$ to be $75\%$ for both photodiodes (Eq.~\ref{equ:dcv}). Coupling efficiency is limited by Fresnel reflections at the air-glass interfaces of the fibre launch, the focus lens and the photodiode housing. It can in principle be improved by using fibre coupled photodiodes with $\leq$\SI{1}{\decibel} insertion loss, which are currently only commercially available, at greater expense, when custom made. 

In measurements of continuous-variable quantum states, the signal-to-noise ratio (SNR) also effectively degrades the total efficiency. For a given SNR in \si{\dB}, the corresponding measurement efficiency is ${\eta_{\mathrm{SNR}} = (\mathrm{SNR}-1)/\mathrm{SNR}}$ \cite{Tasker2021SiliconLight}. In practice, the negative effect of electronic noise on such measurements can be mitigated by increasing LO power to just below the detector saturation power. The SNR of the detector close to saturation is \SI{9}{\decibel} (at \SI{5}{\mega\hertz}) with a corresponding efficiency of $\eta_{\mathrm{SNR}} = 88\%$. The total efficiency of the detectors for squeezed light detection at \SI{5}{\mega\hertz} is therefore limited to {$\eta_{\mathrm{tot}}=\eta_\mathrm{SNR}\eta_\mathrm{coup}\eta_{QE}=58\%$}. The main limitation here is not linked to the fundamental properties of the detector but is instead due to the poor coupling efficiency of the photodiodes. Employing state-of-the-art fibre-coupled photodiodes, the total detector efficiency at \SI{2}{\micro\meter} could be readily increased to $\eta_{\mathrm{tot}}=73\%$. Although this is still below standard telecoms band efficiencies of $\eta_{\mathrm{tot}}=95\%$ \cite{Hoff2015}, sub-shot-noise measurements are possible with such a device.

By developing a shot-noise limited detector that demonstrates {$58\%$} efficiency at megahertz-speeds, we remove a crucial roadblock to quantum state development in the \SI{2}{\micro\meter} band that will help enable further development of prospective technologies including broadband \SI{2}{\micro\meter} squeezing for quantum-enhanced metrology~\cite{Atkinson2021QuantumLight}, SWIR-based ground-to-satellite quantum-key distribution~\cite{Temporao2008FeasibilityMID-INFRARED} and remote LIDAR sensing~\cite{Yu2014DevelopmentMeasurement,Refaat2018AirborneSensing}. Specifically, the increase in speed from kilohertz to megahertz could help facilitate the implementation of CV quantum information processing~\cite{Hamerly2019Large-ScaleMultiplication} and optical neural networks~\cite{Larsen2019DeterministicState,Asavanant2019GenerationState} which stand to benefit from reduced two-photon absorption in silicon in the \SI{2}{\micro\meter} band. By developing a detector that affords a high signal-to-noise ratio, we also help facilitate higher sensitivity in classical schemes that employ balanced detection to monitor amplitude modulation signals~\cite{Lendl2021Balanced-detectionConfiguration}, reject classical laser noise~\cite{Leleux2001AmmoniaSpectroscopy} or map information from optical frequencies to the RF via coherent heterodyne beating of optical frequency combs~\cite{Cole2018DownsamplingCombs}.

The detector bandwidth was designed to sufficiently suppress amplification of the repetition rate of the laser used in our experiment (\SI{39.5}{\mega\hertz}). Without doing so, the detector is at risk of saturating at a single frequency which in turn can cause non-linear behaviour in other parts of the spectrum, which can skew the quadrature measurement. By observing the AC signal on the oscilloscope, we see that the detector saturates at the repetition rate first suggesting greater suppression may increase the saturation power further. However, this would come at the cost of decreased detector bandwidth and must be considered for each specific application. Another important consideration is that the quadrature measurement efficiency has been proven to be negatively affected for detector bandwidths less than half the repetition rate of the laser \cite{Kumar2012VersatileTomography}. This is a direct result of each pulse running into the previous pulse which mixes the temporal mode of the quadrature measurement across consecutive pulses. One approach to increase the bandwidth of shot-noise limited performance is to miniaturise the detector's electronic and optoelectronic components to curtail overall capacitance~\cite{Tasker2021SiliconLight}.  \textit{GeSn} integrated detectors for integrated photonics platforms are also being developed that are sensitive to the SWIR and promise a considerable increase in speed~\cite{Tan2019IntegratingBand}.

The best CMRR achieved for \SI{1550}{\nano\meter} homodyne detectors is \SI{75.2}{\decibel} and requires fine-tuning circuitry to account for slight differences in transit time and responsivity between the two photodiodes~\cite{Jin2015BalancedPhotodiodes}. Without this additional path tuning, small variations in transit time make it difficult to improve the CMRR beyond $~$\SI{60}{\decibel}. The path length difference between each arm was fine-tuned to maximise rejection inferring \SI{48}{\decibel} to be limited by such variations.

\section{Conclusion}

We have presented a shot-noise limited homodyne detector designed for quantum measurement of pulsed light in the \SI{2}{\micro\meter} band. Using this detector, we measure megahertz vacuum fluctuations and verified the detector to be shot-noise limited, indicating its efficacy for detecting quantum light and for characterising quantum light fields. Through judicious design, we prove it possible to overcome the sub-optimal electrical characteristics of the available photodiodes responsive in the \SI{2}{\micro\meter} band to provide sufficient shot-noise clearance at speeds that cannot be detected by previous demonstrations of shot-noise limited homodyne detectors. We comment on the efficiency of the detector in relation to applications of squeezed light. We note the main limitation on device efficiency to be coupling efficiency, not the fundamental speed or dark current of the extended InGaAs photodiodes. The advent of such a detector opens the door to the measurement and verification of exotic quantum states in the \SI{2}{\micro\meter} band and is of sufficiency efficacy to employ in future cutting-edge quantum noise limited and quantum-enhanced metrology schemes. Additionally, this detector can also provide a route towards increased signal-to-noise ratio of balanced detectors utilised in schemes to measure weak modulation signals.

\section{ACKNOWLEDGEMENTS \& FUNDING}

The authors would like to thank helpful and informative discussions with J. Frazer, G. Ferranti, D. Payne and E. J. Allen. 

This work was supported by the Centre for Nanoscience and Quantum Information (NSQI), EPSRC UK Quantum Technology Hub QUANTIC EP/M01326X/1. J.B. acknowledges support from EPSRC Quantum Engineering Centre for Doctoral Training EP/LO15730/1. J.F.T. acknowledges studentship support from EPSRC. J.W.S acknowledges support from the Leverhulme Trust ECF-2018-276 and UKRI Future Leaders Fellowship MR/T041773/1. J.C.F.M. acknowledges support from an ERC starting grant ERC-2018-STG 803665.

\section{Data Availability}

Data underlying the results presented in this paper, along with Altium board files for the detector, are available in reference~\cite{Biele2021Jake-Biele/Homodyne_detection}.

\bibliographystyle{ieeetr}

\onecolumngrid

\end{document}